 \documentclass[smallabstract,smallcaptions]{dccpaper}

\usepackage{epsfig}
\usepackage{citesort}
\usepackage{amsmath}
\usepackage{amssymb}
\usepackage{color}
\usepackage{url}

\newlength{\figurewidth}
\newlength{\smallfigurewidth}

\begin{document}

\title
{\large
\textbf{Compressed Domain Prior-Guided Video Super-Resolution for Cloud Gaming Content}
}

\author{%
Qizhe Wang$^{\ast}$, 
Qian Yin$^{\ast}$,
Zhimeng Huang$^{\ast}$,
Weijia Jiang$^{\dag}$,
Yi Su$^{\dag}$,  \\Siwei Ma$^{\ast}$ and 
Jiaqi Zhang$^{\ast}$\thanks{Correspondence to: J. Zhang(jqzhang@pku.edu.cn)}\\[0.5em]
{\small\begin{minipage}{\linewidth}\begin{center}
\begin{tabular}{ccc}
$^{\ast}$National Engineering Research Center of Visual Technology,\\
School of Computer Science, Peking University, China \\
$^{\dag}$Migu Interactive Entertainment Co., Ltd, China \\
\url{{qzwang, yinqian_xixi}@stu.pku.edu.cn}, %\url{yinqian_xixi@stu.pku.edu.cn}, 
\url{{zmhuang, swma, jqzhang}@pku.edu.cn},\\
%\url{cmjia@pku.edu.cn}, 
\url{{jiangweijia, suyi}@migu.chinamobile.com}
%, \url{swma@pku.edu.cn}
\end{tabular}\\[0.5em]
\end{center}\end{minipage}}
}

\maketitle
\thispagestyle{empty}

\begin{abstract}
Cloud gaming is an advanced form of Internet service that necessitates local terminals to decode within limited resources and time latency. 
Super-Resolution (SR) techniques are often employed on these terminals as an efficient way to reduce the required bit-rate bandwidth for cloud gaming. 
However, insufficient attention has been paid to SR of compressed game video content. 
Most SR networks amplify block artifacts and ringing effects in decoded frames while ignoring edge details of game content, leading to unsatisfactory reconstruction results. 
In this paper, we propose a novel lightweight network called Coding Prior-Guided Super-Resolution (CPGSR) to address the SR challenges in compressed game video content. 
First, we design a Compressed Domain Guided Block (CDGB) to extract features of different depths from coding priors, which are subsequently integrated with features from the U-net backbone. 
Then, a series of re-parameterization blocks are utilized for reconstruction. 
Ultimately, inspired by the quantization in video coding, we propose a partitioned focal frequency loss to effectively guide the model's focus on preserving high-frequency information.
Extensive experiments demonstrate the advancement of our approach.
% Cloud gaming is an advanced form of Internet service that necessitates local tminals to decode within limited resources and time latency, while super-resolution (SR) is often employed on these terminals as an efficient technique to reduce the required bitrate bandwidth for cloud gaming. However, little attention has been paid to SR of compressed video and game content. Most SR networks amplify block artifacts and ringing effects in decoded frames while ignoring edge details of game content, leading to unsatisfactory reconstruction results. In this paper, we propose a novel lightweight network called Coding Prior-Guided Super-Resolution (CPGSR) to solve the SR problem of compressed game content. First, we design a Compressed Domain Guided Block (CDGB) to extract features  of different depths from coding priors, which are subsequently integrated with features from the U-net backbone. Then, a series of re-parameterization blocks are utilized for reconstruction. Ultimately, inspired by the quantization in video coding, we propose a partitioned focal frequency loss to effectively guide the model's attention towards preserving high-frequency information. Extensive experiments demonstrate the advancement of our approach.
\end{abstract}

\Section{Introduction}
With the development of cloud computing and network transmission, cloud gaming has become one of the emerging Internet industries. In the cloud gaming system, cloud nodes execute games according to instructions of the user's local terminal, and transmit the game content to the user through the network. Due to bandwidth limitations and unstable wireless transmission, the cloud needs to compress game content to reduce the transmission bitrate and ensure smooth gaming. Over the past few years, the coding of cloud gaming content has received increasing attention, in which JVET AHG15~\cite{JVETahg15} has made a great contribution. Among the various coding methods, an effective approach is to reduce the resolution of game content, and then implement decoding and Super-Resolution (SR) of the game videos by the user's local terminal within limited resources and time constraints. SR aims to reconstruct High Resolution (HR) images or videos from Low Resolution (LR), which is widely used in applications where high-quality images are necessary. Since the pioneering work of Dong et al.~\cite{dong2014learning}, numerous neural networks have been developed to tackle this challenge. In general, deeper and more complex network structures can provide better performance, but the high computational cost also limits the application scenarios of these methods. In recent years, many efficient SR studies have emerged \cite{ignatov2021real,li2023ntire,conde2024real,khan2022ntire,ignatov2022efficient}, most of which only consider how to continuously infer high-frequency details. However, they ignore the impact of compression distortion and specific characteristics of video content on SR techniques. Therefore, most methods cannot effectively handle compressed game content, which exacerbates the blocking and ringing artifacts in decoded frames. Moreover, they fail to address the intricate edges inherent in game content, ultimately leading to unsatisfactory reconstruction results. 

\begin{figure*}[t]
\centering
\includegraphics[width=3.3in]{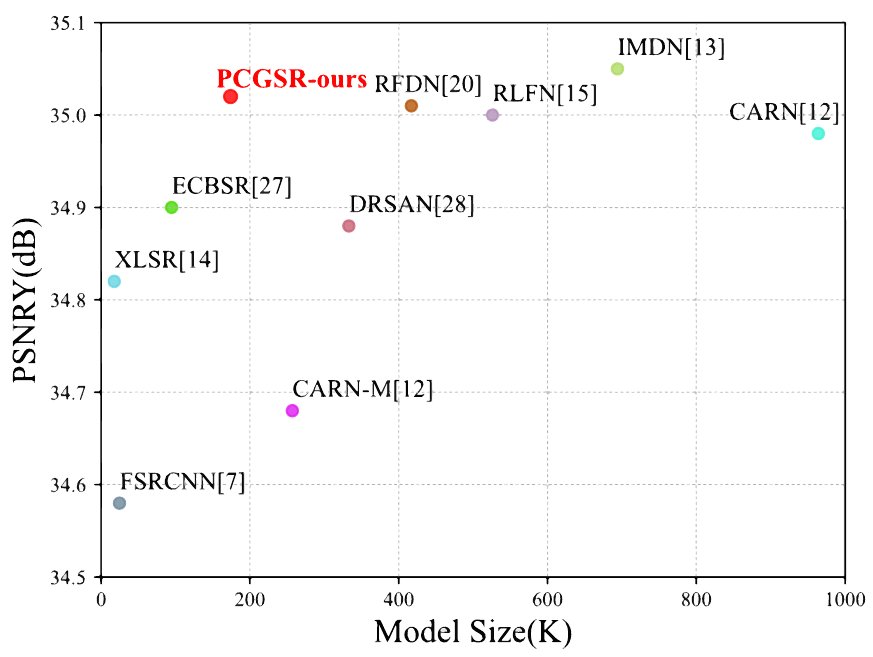}
\caption{
Performance comparison of our CPGSR with nine other lightweight models on the proposed dataset, which is evaluated in terms of PSNRY and model parameter size.}
\label{bubble}
\end{figure*}

To overcome the above-mentioned problems, we propose a Coding Prior-Guided Super-Resolution (CPGSR) specifically designed for the SR of the compressed game screen contents. As shown in Fig.~\ref{bubble}, the proposed method achieves an excellent trade-off between performance and computing resources, which supports cloud gaming content with refresh rates exceeding 60Hz. The contributions of this paper can be summarized as follows:
\begin{itemize}
    \item We present a novel lightweight SR network that utilizes coding priors for compressed game content. The network includes three parts, wherein a Compressed Domain Guided Block (CDGB) for coding priors information extraction, a U-net backbone for deep feature extraction and fusion, and a series of re-parameterization blocks for reconstruction.
    \item We propose a partitioned focal frequency loss to guide the model in recovering high-frequency information lost during video compression.
    \item We built a new compressed cloud gaming content dataset based on the Versatile Video Coding (VVC) standard. The proposed dataset has a variety of coding priors as well as compressed LR data with the associated pristine HR counterpart, aiming to promote the research in the field of compressed cloud gaming content enhancement.
\end{itemize}

\Section{Related Work}
\SubSection{Efficient Super-Resolution Methods}
Lightweight is a critical consideration for SR models, and many approaches have been proposed to improve their efficiency. Dong et al. \cite{dong2014learning} first applied deep learning to single-image SR, involving three convolutional layers, then FSRCNN \cite{dong2016accelerating} was composed to accelerate the network by adopting smaller size of convolution kernels and postponing the upscaling layer at the end of the network. After that, ESPCN \cite{shi2016real} introduced a sub-pixel convolutional layer, which is now widely used in numerous SR network \cite{yu2023dipnet,guo2023asconvsr,bilecen2023bicubic++}. In order to further improve efficiency, CARN \cite{ahn2018fast} used group convolutions and a cascading mechanism but impaired the performance, while IMDN \cite{10.1145/3343031.3351084} applied the three step-distillation to extract features and a slice operation to divide the extracted feature, but brought inflexibility. Later on, with the increasing attention to efficient super-resolution, many efficient SR studies have emerged. XLSR \cite{ayazoglu2021extremely} used clipped ReLU at the last layer of the network and achieved a great balance between reconstruction quality and runtime. And RLFN \cite{9857179} utilized three convolutional layers for residual local feature learning to simplify feature aggregation. In addition, the proposal of re-parameterization \cite{ding2021repvgg}  allowed complex blocks to be used to train the network, while the so-called `RepBlocks' could be reduced to simple 3$\times$3 convolutions during the inference phase. 
Inspired by this work, numerous efficient re-parameterization blocks have been proposed to solve problems in different scenarios \cite{zhang2021edge,zhang2022lightweight,wang2024repvit1}.
This technique has become state-of-the art in efficient SR \cite{conde2024real}. As shown in Fig.~\ref{bubble}, compared with the models mentioned above, our method achieves better trade-off between performance and run time.

\SubSection{Utilization of Coding Priors in Deep Networks}
In compressed domain processing, coding priors usually serve as side information for deep networks to enhance the quality of reconstructed frames. 
In~\cite{lin2019partition}, the close relationship between partition maps and blocking artifacts is utilized as a guide for the network to optimize image details. Chen et al. \cite{wang2023compressed,chen2021compressed} proposed an SR method for compressed videos, which jointly exploited the coding priors and deep priors. However, these methods are too complex to support cloud gaming and 
do not consider the characteristics of coding priors when extracting features.
% the characteristics of coding priors are not considered when extracting features.

\begin{figure*}[ht]
\centering
\includegraphics[width=4.8in]{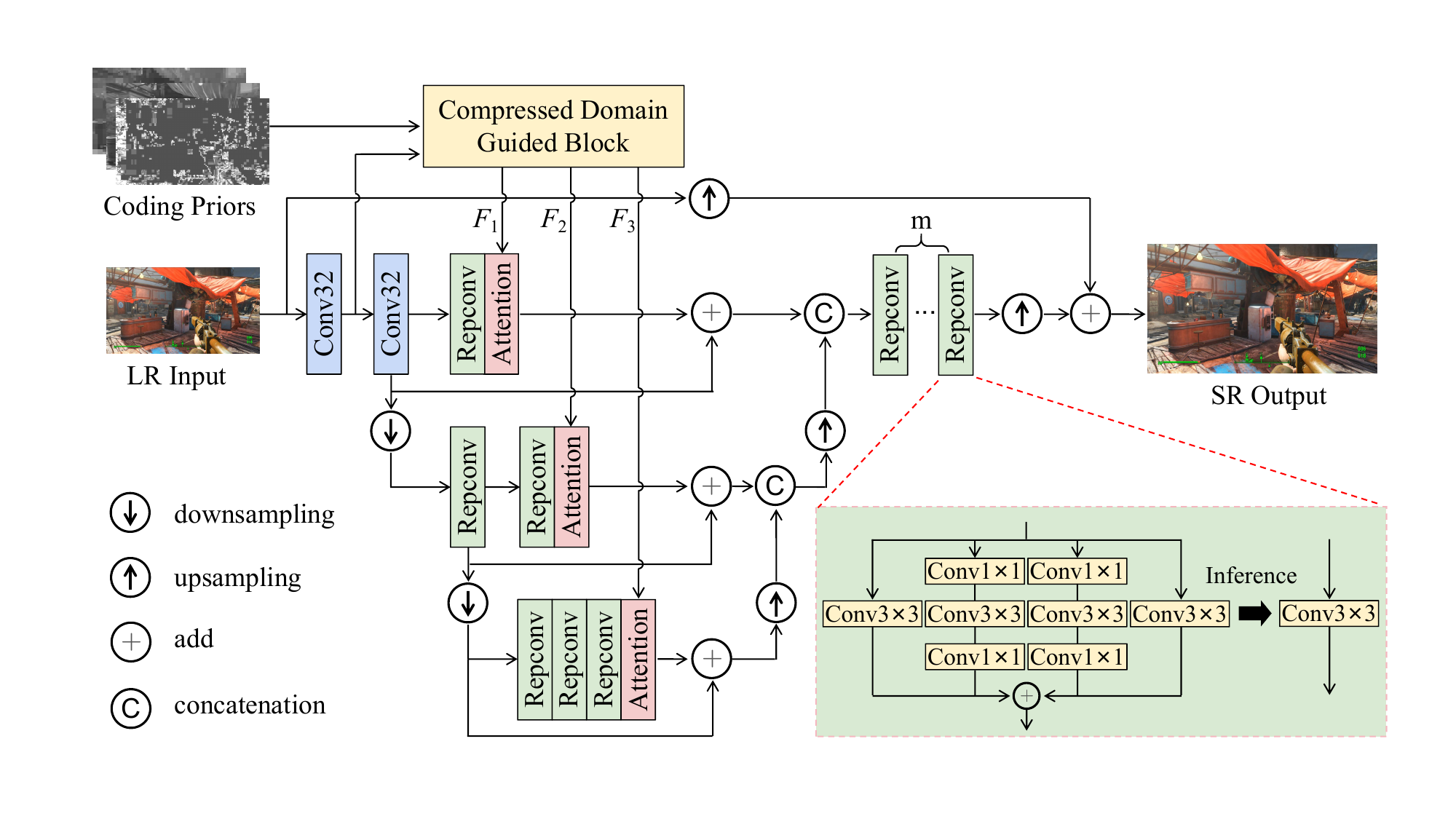}
\caption{The architecture of Coding Prior-Guided Super-Resolution (CPGSR). The proposed network is composed of three parts: Compressed Domain Guided Block (CDGB), mutual information processing and reconstruction. The CDGB part extracts compression features from coding priors. Then the compression features of different depths are fed into the U-net backbone and fused with the input features. Finally, m Repconv blocks are used for the reconstruction of the frame.}
\label{architecture}
\end{figure*}

\Section{Proposed Method}
The architecture of our proposed model is shown in Fig.~\ref{architecture}. 
It consists of three parts: Compressed Domain Guided Block (CDGB), mutual information processing and reconstruction. We first use a 32-channel 3$\times$3 convolution layer to encode the LR input into a high-dimensional feature space. Subsequently, the LR features are fed into both U-net and CDGB architectures. In CDGB architecture, coding priors and LR features are integrated to endow U-net with informative side information at different depths. Then the side information is fused with LR features and finally a series of Repconv blocks is used for reconstruction.

\SubSection{The Compressed Domain Guided Block}
Although video coding introduces compression artifacts, the coding priors can serve as valuable side information for video processing. The prediction signal and the coding residual implicitly reflect the relative changes in video content. The Quantization Parameter (QP) determines the quantization step size. The partition map provides valuable insights into the general shape of object, intra-frame texture complexity, and the distribution of blocking artifacts caused by coding. Therefore, we propose a CDGB module to extract coding prior information for quality enhancement of reconstructed images.

The full CDGB structure details are shown in Fig.~\ref{comblock}. First, we use a 32-channel 3$\times$3 convolution layer to extract features from a series of coding priors, including prediction signals, QP and residuals. Then, we derive a pair of affine transformation parameters based on the characteristics of the input frame and the coding priors, which can be described as follows:
\begin{equation}
F_{0}  = \gamma  \odot  F_{in}  + \beta,
\end{equation}
where $F_{in}$ represent LR features and $\odot$ is the element-wise multiplication. The transformation parameters $\gamma$ and $\beta$ are generated by jointly considering the coding priors features and LR features. Subsequently, three Repconv blocks (i.e., $f_{Rep}$) and two Pixel-Adaptive Convolution (PAC) blocks (i.e., $f_{PAC}$) are employed to extract informative side information at different depths, which can be described as follows:
\begin{figure*}[ht]
\centering
\includegraphics[width=4.8in]{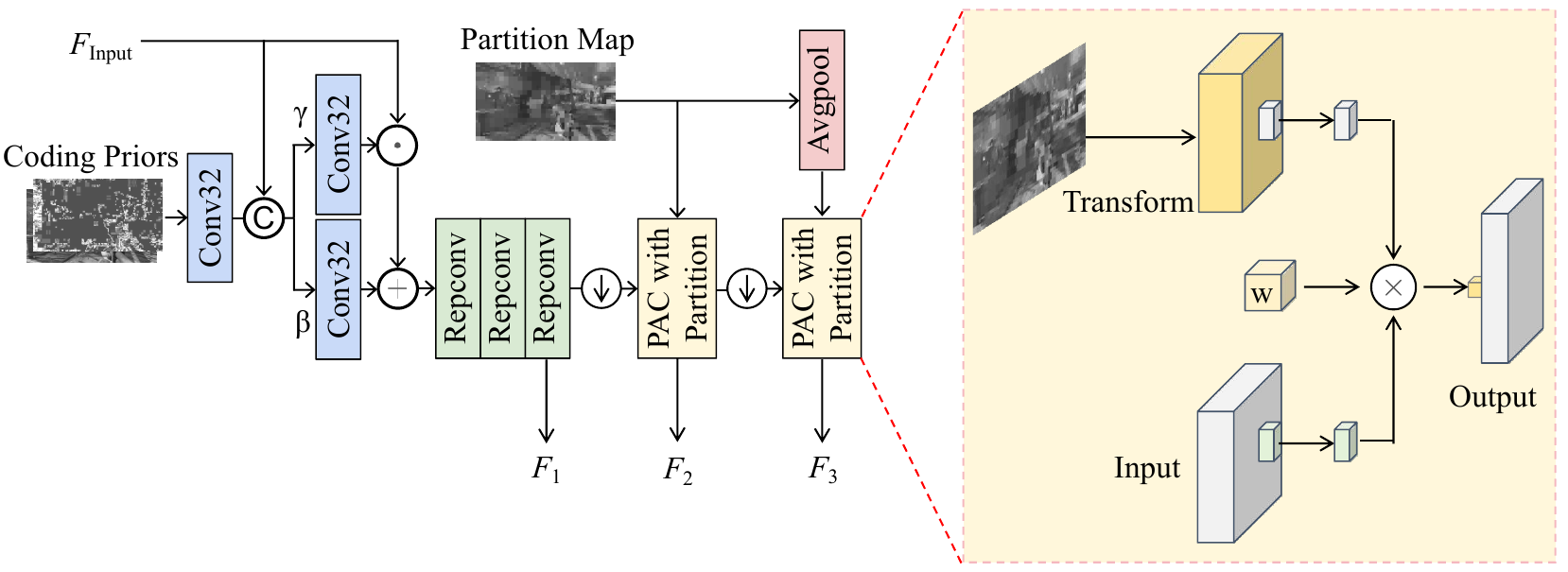}
\caption{
The architecture of CDGB. A pair of affine transformation parameters is derived based on both the LR features and the coding priors, then compression features of different depths are generated by Repconvs and two PAC blocks. In the  PAC block, the partition map is transformed into Pixel-Adaptive weights and participates in the convolution process.}
\label{comblock}\vspace{-0.1in}
\end{figure*}
\begin{equation}
\begin{aligned}
F_{1}  &= f_{Rep}(F_{0}),
\\F_{2}  &= f_{PAC}(f_{\downarrow }(F_{1}), Part),
\\F_{3}  &= f_{PAC}(f_{\downarrow }(F_{2}), AvgPool(Part,  2)),
\end{aligned}
\end{equation}
where $f_{\downarrow}$ represents a downsampling operation achieved through the implementation of a 3$\times$3 convolutional block with a stride value of 2. $Part$ is the partition map of the input frame, and $AvgPool()$ is the AveragePool function. $F_{1}, F_{2}, F_{3}$ represent compressed domain information at different depths.

\begin{figure}[t]
\begin{center}
\begin{tabular}{cc}
\includegraphics[width=2in]{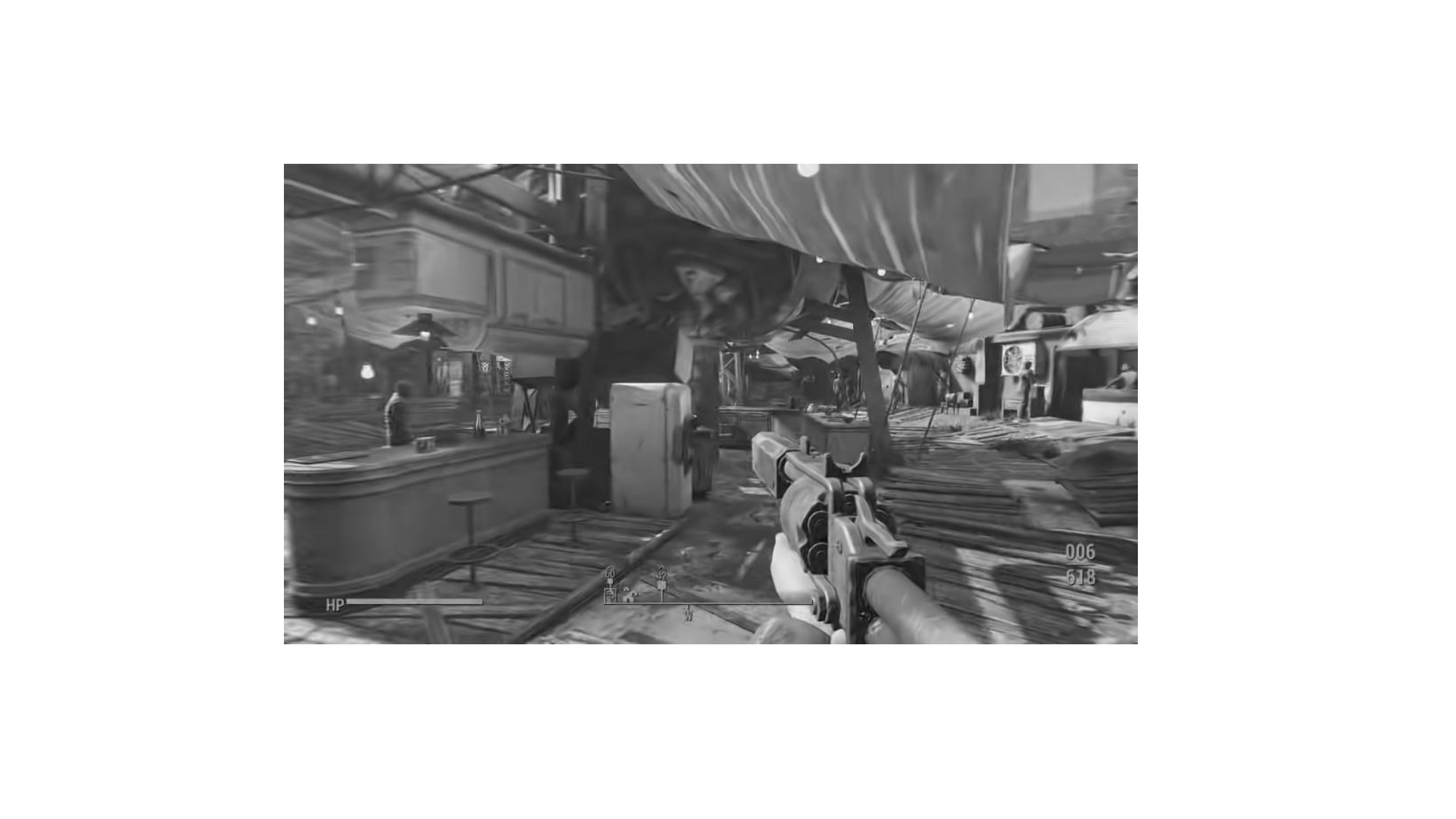} &
\includegraphics[width=2in]{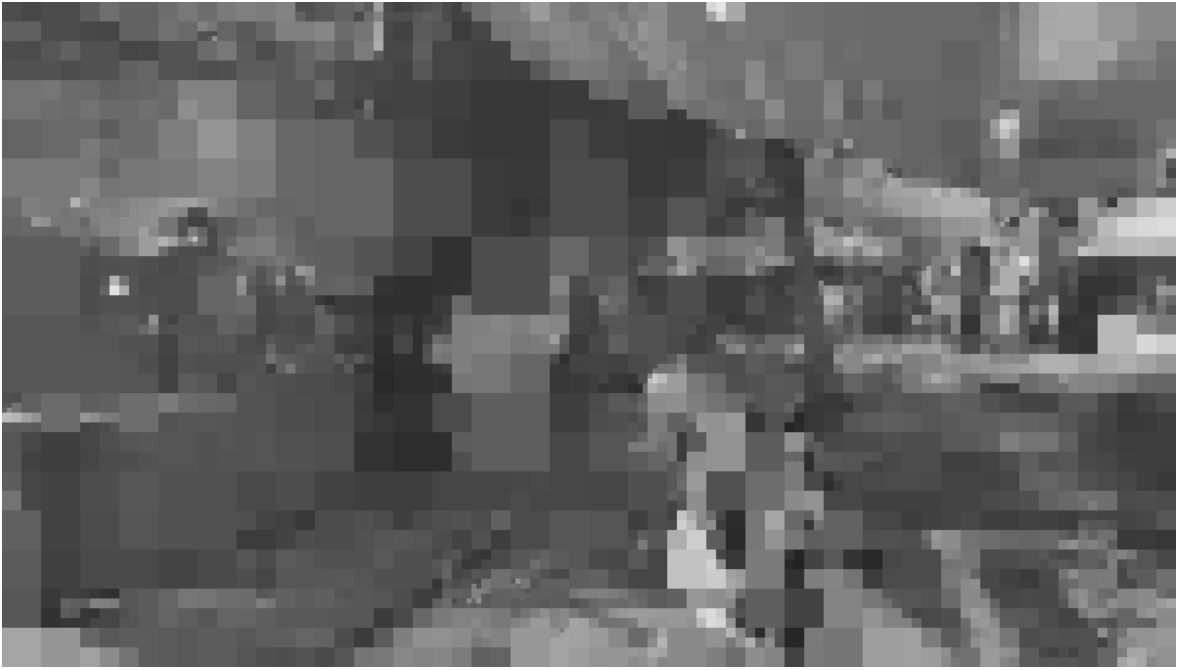}\\
{\small (a)} & {\small (b)}
\end{tabular}
\end{center}\vspace{-0.3in}
\caption{\label{fig:Partition}%
(a) The original decoded frame in grayscale (b) The partition map of the frame}
\end{figure}

The structure of PAC with partition is shown in Fig.~\ref{comblock}.
% Convolutions are the fundamental building blocks of CNNs. The fact that their weights are spatially shared is one of the main reasons for their widespread use, but it is also a major limitation, as it makes convolutions content-agnostic.
Convolutions, as the basic building block of CNNs, are widely used because their weights are shared spatially. But this is also a major limitation because it makes convolutions independent of content. To tackle this problem, Su et al. \cite{su2019pixel} proposed the PAC operation, in which the filter weights are multiplied with a spatially varying kernel that depends on learnable local pixel features. Based on this premise, we transform the partition map into spatially varying kernels, enabling the model to capture the texture structure information of the current frame. As shown in Fig.~\ref{fig:Partition}, the partition map delivers useful hints regarding the object shapes and complexity of textures. Formally, the PAC with partition can be formulated as:
\begin{equation}
    v_{i}^{'} = \sum_{j\in \Omega \left(i\right)}F(Pa_{i}, Pa_{j})W\left [ p_{i}-p_{j} \right ] v_{j}+b.
\end{equation}
Herein, $p_{i} \in \left ( x_{i} , y_{i} \right )^\top$ are pixel coordinates. $\Omega \left(*\right)$ defines a $s\times s$ convolution window. The image features $v = \left(v_{1},v_{2},...,v_{n}\right), v_{i}\in\mathbb{R} ^{c}$ over n pixels and c channels. The filter weights $W \in\mathbb{R} ^{c'\times c\times s\times s }$ and $b \in \mathbb{R}^{c'}$ denotes bias. In addition, $F(Pa_{i}, Pa_{j})$ is the function that generates the weights of the relative position convolution kernel based on the partition map information, and the $Pa_{i}$ is the partition map value at position i. $F(Pa_{i}, Pa_{j})$ can be written as:
\begin{equation}
   F(Pa_{i}, Pa_{j}) = clamp(-1*Threshold*(Pa_{j}-Pa_{i})^2 + 1, 0),
\end{equation}
where the threshold is set as 16 and $clamp()$ set the negative weight to 0. By this means, the partition map assigns a convolution weight to each pixel, allowing the filters to be modified by the adapting kernel $K$ differently across pixel locations.

\SubSection{The U-net backbone}
In the U-net backbone, the input features undergo two downsampling operations implemented by 3$\times$3 convolution layer with a stride of 2, which are then integrated with side information in the Attention block. The upsampling operation consists of a 1$\times$1 convolution followed by pixel shuffle. In the Attention block, we utilize feature gating to modulate the input features and employ channel attention to extract global information. Following this, two fully-connected layers with GELU activation functions are applied. 

Furthermore, we employ a re-parameterization block, termed `Repconv Block', as shown in Fig.~\ref{architecture}. During the training phase, the input features are processed by two 3$\times$3 convolutional layers and one convolutional layer consisting of two 1$\times$1 convolutions and one 3$\times$3 convolutions, and the training results of the different convolutions will be summed up at the output, which will be transformed into a simple 3$\times$3 convolution in the inference phase.

\SubSection{The Partitioned Focal Frequency Loss}
To alleviate the gaps between the real and generated images in the frequency domain, Jiang et al. propose a novel focal frequency loss \cite{jiang2021focal}, which allows a model to adaptively focus on frequency components that are hard to synthesize by down-weighting the easy ones. Similarly, analogous challenges arise in the compressed domain when comparing compressed frames with their original counterparts. In the compressed video, the loss of high-frequency information primarily arises from the transformation and quantization operations performed on Coding Units (CUs). Therefore, the Partitioned Focal Frequency Loss (PFFL) is proposed to solve the problem. Firstly, the SR outputs and HRs $P_{SR}, P_{HR} \in {(3,H, W)}$ are partitioned into CUs of 32$\times$32 pixels. These CUs are then subjected to a fast Fourier transform, resulting in frequency domains $Fre_{cSR}, Fre_{cHR} \in {(\frac{3\times H\times W}{32\times 32}, 32, 32)}$. Here $H, W$ represent the height and width of each frame, respectively. Then a spectrum weight matrix is constructed, characterized by a locked gradient. Consequently, it functions as the weighting factor for each frequency component. The weight matrix can be written as:
\begin{equation}
   w(x, y, z)=\left|Fre_{cSR}(x, y, z)-Fre_{cSR}(x, y, z)\right|^{\alpha},
\end{equation}
where $\alpha$ is the scaling factor for flexibility ($\alpha$ = 1 in our experiments), and $(x, y, z)$ is the spectrum coordinate. We further normalize the matrix values into the range [0, 1]. Finally, the full form of PFFL can be written as :
\begin{equation}
   \mathrm{PFFL}=\frac{1}{\sqrt{H W}} \sum_{u=0}^{31} \sum_{v=0}^{31} \sum_{w=0}^{ \frac{H \times W \times 3}{1024}-1} w(x, y, z)\left|Fre_{cSR}(x, y, z)-Fre_{cHR}(x, y, z)\right|^{2}.
\end{equation}

\Section{Experiments}
In this section, we describe in detail the performance of our method for SR tasks and the ablation experiments for each module.
\SubSection{Datasets}
Herein, we collect 10 different types of game video sequences with the resolution 1080p from the public domain, each containing 900 frames. The corresponding downsampled LR version is generated using the Bicubic method with a factor 2. Following this, the LR dataset is compressed using an opensource software VVenC-1.12.0 \cite{VVenC} that complies with VVC video coding standard. We further extract four types of coding priors embedded in the bitstream, including frame level QP, prediction signal, coding residuals, and partition map.

\SubSection{Implement Details}
During training, the initial learning rate is set to 5e-4, and the training stop after 200 epochs with early-stopping mechanism. The optimizer used is the Adam optimizer with $\beta$1 of 0.9 and $\beta$2 of 0.999. In addition, we employ randomly rotating 90$^{\circ}$, 180$^{\circ}$, 270$^{\circ}$, vertical flip and horizontal flip for data augmentation. In the final model, the output channel is set to 32 for all Repconv blocks and the depth multiplier of the block is set to 2. The loss function we used can be written as:
\begin{equation}
   \mathrm{Loss}=\alpha \times \mathrm{L1} +  \beta \times \mathrm{PFFL},
\end{equation}
where $\alpha, \beta$ are set to 0.9, 0.1 in the implementation.
Since the format of the decoded frames is YUV420p, we first convert the input frame to RGB format and re-convert the output frame to YUV420p for evaluations. We mainly use the commonly used evaluation metrics, including Peak Signal-to-Noise Ratio (PSNR) on Y,U,V channels and Structural Similarity (SSIM) on Y channel.
\begin{table}[!ht]
    \centering
    \small 
    \caption{\label{compare}%
    Average PSNR and SSIM comparison with other lightweight models.}
    \begin{tabular}{|c|c|c|c|c|c|}
    \hline
        Method & Params(K) & PSNRY(dB) & PSNRU(dB) & PSNRV(dB) & SSIM \\ \hline
        FSRCNN \cite{dong2016accelerating}& 24.70 & 34.58 & 39.85 & 39.85 & 0.8199  \\ 
        CARN-M \cite{ahn2018fast}& 257.00 & 34.68 & 39.90 & 39.90 & 0.8218 \\ 
        ECBSR \cite{ecbsr}& 94.70 & 34.90 & 40.51 & 40.50 & 0.8305 \\ 
        XLSR \cite{ayazoglu2021extremely}& 17.60 & 34.82 & 40.48 & 40.49 & 0.8335 \\ 
        DRSAN \cite{9645163}& 333.00 & 34.88 & 40.59 & 40.59 & 0.8308 \\
        IMDN \cite{10.1145/3343031.3351084}& 694.00 & \textbf{35.05} & 40.86 & 40.88 & 0.8350 \\ 
        CARN \cite{ahn2018fast}& 964.00 & 34.98 & 40.73 & 40.73 & \textbf{0.8356}  \\ 
        RFDN \cite{liu2020residual}& 417.00 & 35.01 & 40.60  & 40.61 & 0.8333 \\ 
        RLFN \cite{9857179}& 526.00 & 35.00 & 40.80 & 40.79 & 0.8336 \\ \hline
        \textbf{PCGSR-ours}& 174.00 & 35.02 & \textbf{41.03} & \textbf{41.03} & 0.8352 \\ \hline
    \end{tabular}
\end{table}

\SubSection{Comparison with Lightweight Models}
We compare our model with nine state-of-the-art lightweight SR models, including CARN \cite{ahn2018fast}, RFDN \cite{liu2020residual}, RLFN \cite{9857179}, ECBSR \cite{ecbsr}, etc., as shown in Table~\ref{compare}. Generally speaking, our approach achieves a trade-off between performance and computational resources. Compared to smaller real-time SR methods, our approach achieves significant improvements in PSNR and SSIM values through a moderate increase in parameter count, while ensuring support for cloud game content presentation at refresh rates above 60Hz. Furthermore, our approach delivers comparable performance to more complex SR networks,  achieving over 50$\%$ reduction in parameter count. The experimental results demonstrate that coding priors effectively guide the reconstruction of frames, particularly enhancing chromaticity information where our method outperforms others in terms of PSNRU and PSNRV.
\SubSection{Ablation Study}
We further conduct ablation studies to better understand and evaluate each component in the proposed PCGSR. To ensure fairness, all experiences are conducted under the same setting. The experimental results are presented in the Table~\ref{abalation}. Firstly, the incorporation of coding priors is crucial for enhancing model performance. By leveraging coding priors, the network acquires guidance information from different depths, facilitating frame reconstruction to capture texture details. Then, we substitute PAC and Attention blocks with 3$\times$3 convolutional layers, respectively. The experimental results show that the partition map provides valuable guidance for generating side information. The PAC block mitigates block artifacts by adaptively adjusting weights based on the frame structure and Attention blocks provide superior fusion options for the model.
Lastly, PFFL enables direct focus on high-frequency loss information, significantly augmenting chromaticity details in the reconstructed frame. Moreover, we assess the impact of different numbers of Repconv blocks in the reconstruction stage on the model's performance, and the experimental results are shown in Table~\ref{numofm}. In general, as the depth of the network increases, 
\begin{table}[!ht]
    \centering
    \small 
    \caption{\label{abalation}%
    Average PSNR and SSIM comparison of different settings of our model.}
    \begin{tabular}{|c|c|c|c|c|c|}
    \hline
        Method & PSNRY(dB) & PSNRU(dB) & PSNRV(dB) & SSIM \\ \hline
        w/o CDGB & 34.92 & 40.80 & 40.78 & 0.8333 \\
        w/o PAC & 34.99 & 40.96 & 40.95 & 0.8346 \\
        w/o Attention & 35.00 & 40.90 & 40.90 & 0.8343 \\
        w/o PFFL & 35.01 & 40.92 & 40.91 & 0.8345 \\ \hline
        \textbf{Full} & \textbf{35.02} & \textbf{41.03} & \textbf{41.03} & \textbf{0.8352} \\ \hline
    \end{tabular}
\end{table}
\begin{table}[!ht]
    \centering
    \small 
    \caption{\label{numofm}%
    Average PSNR and SSIM comparison of different number of m.}
    \begin{tabular}{|c|c|c|c|c|c|}
    \hline
        Number of m & Params(K) & PSNRY(dB) & PSNRU(dB) & PSNRV(dB) & SSIM \\ \hline
        3 & 170.00 & 34.94 & 40.93 & 40.94 & 0.8343 \\ \hline
        4 & 172.00 & 34.98 & 41.00 & 40.99 & 0.8347 \\ \hline
        5 & 174.00 & \textbf{35.02} & \textbf{41.03} & \textbf{41.03} & \textbf{0.8352} \\ \hline
        6 & 176.00 & 34.98 & 40.94 & 40.91 & 0.8341 \\ \hline
    \end{tabular}
\end{table}there is an improvement in model performance; however, excessive stacking of convolutional layers may lead to a decline in performance. In this study, for the final stage of the SR task, optimal performance is achieved when m=5.

\Section{Conclusion}
In this paper, we have proposed a novel lightweight SR network leveraging the coding priors to achieve efficient SR for compressed cloud gaming content. We have also introduced a partitioned focal frequency loss to guide the model in recovering high-frequency information lost during video compression. Additionally, a corresponding game content dataset is created to facilitate the research of cloud gaming content processing. Extensive experimental results show the superior performance of our proposed scheme in terms of restoration quality and computational complexity.

\Section{Acknowledgement}
This work was supported in part by NSFC No. 62025101, in part by Fundamental Research Funds for the Central Universities, in part by the Postdoctoral Fellowship Program of CPSF under Grant Number GZC20230059 and GZC20240035, in part by R24115SG MIGU-PKU META VISION TECHNOLOGY INNOVATION LAB, in part by New Cornerstone Science Foundation through the XPLORER PRIZE.

\Section{References}
\bibliographystyle{IEEEbib}
\bibliography{refs}

\end{document}